\pdfoutput=1

\documentclass[11pt]{article}

\usepackage[final]{acl}

\usepackage{times}
\usepackage{latexsym}

\usepackage[T1]{fontenc}

\usepackage[utf8]{inputenc}

\usepackage{microtype}

\usepackage{inconsolata}

\usepackage{graphicx}

\usepackage{microtype}
\usepackage{hyperref}
\usepackage{url}
\usepackage{booktabs}

\usepackage{enumitem}

\usepackage{framed}
\usepackage{lmodern}  
\usepackage{amsmath}
\usepackage{float}
\usepackage{anyfontsize}
\usepackage[inkscapelatex=false]{svg}
\usepackage{colortbl}
\usepackage{array}
\usepackage{mdframed}
\usepackage{tcolorbox}
\usepackage{wrapfig}
\usepackage{adjustbox}
\usepackage{tabularx}

\usepackage{xspace}
\usepackage{colortbl}
\usepackage{xspace}
\usepackage{enumitem}
\usepackage{soul}
\usepackage{dirtytalk}
\usepackage{epigraph} 
\usepackage{multirow}
\usepackage{graphicx}
\usepackage{wrapfig}

\newcommand{\newparagraph}[1]{{\vspace{3pt}{\noindent\textbf{#1}}}}

\newcommand{\system}{{\textsc{ValueCompass}}\xspace}

\definecolor{hua-red}{HTML}{DE4A4D}
\definecolor{HUA-RED}{HTML}{DE4A4D}
\definecolor{hua-blue}{HTML}{1271CA}
\definecolor{HUA-BLUE}{HTML}{1271CA}
\definecolor{hua-yellow}{HTML}{F9AB10}
\definecolor{HUA-YELLOW}{HTML}{F9AB10}
\definecolor{hua-green}{HTML}{2E9B42}
\definecolor{hua-purple}{HTML}{533283}

\definecolor{darkblue}{rgb}{0, 0, 0.5}
\hypersetup{colorlinks=true, citecolor=darkblue, linkcolor=darkblue, urlcolor=darkblue}

\title{\system: A Framework for Measuring Contextual Value Alignment Between Human and LLMs}

\usepackage{comment}
\usepackage{multicol}
\usepackage{multirow}
\usepackage{epsfig}
\usepackage{pifont}
\usepackage{dsfont}
\usepackage{makecell}
\usepackage{adjustbox}
\usepackage{txfonts}
\usepackage{xcolor}
\usepackage{soul}
\usepackage{caption}
\usepackage{subcaption}

\author{
Hua Shen\textsuperscript{${\varheartsuit \vardiamondsuit}$}
\quad Tiffany Knearem\textsuperscript{${\diamondsuit}$}
\quad Reshmi Ghosh\textsuperscript{${\dagger}$}
\quad Yu-Ju Yang\textsuperscript{${\heartsuit}$} \\
\quad {\bf Nicholas Clark}\textsuperscript{${\vardiamondsuit}$}
\quad {\bf Yun Huang} \textsuperscript{${\heartsuit}$}
\quad {\bf Tanu Mitra}\textsuperscript{${\vardiamondsuit}$}
\\ 
\textsuperscript{${\varheartsuit}$} NYU Shanghai, New York University, 
\textsuperscript{${\vardiamondsuit}$}University of Washington, \\
\textsuperscript{${\diamondsuit}$}MBZUAI,
\textsuperscript{${\dagger}$}Microsoft, 
\textsuperscript{${\heartsuit}$}UIUC\\
    {\tt huashen@nyu.edu},
    {\tt Tiffany.Knearem@mbzuai.ac.ae},
    {\tt reshmighosh@microsoft.com},
    \\
    {\tt nclark4,tmitra@uw.edu}, 
    {\tt yuju2,yunhuang@illinois.edu},
  }

\begin{document}


\maketitle

\begin{abstract}

As AI advances, aligning it with diverse human and societal values grows critical. But how do we define these values and measure AI’s adherence to them?
We present \system, a framework grounded in psychological theories, to assess human-AI alignment. Applying it to five diverse LLMs and 112 humans from seven countries across four scenarios—collaborative writing, education, public sectors, and healthcare—we uncover key misalignments. For example, humans prioritize national security, while LLMs often reject it. Values also shift across contexts, demanding scenario-specific alignment strategies.
This work advances AI design by mapping how systems can better reflect societal ethics\footnote{Data and code are released on Github: \href{https://github.com/huashen218/value_action_gap}{https://github.com/huashen218/value\_action\_gap}}.



\end{abstract}

\section{Introduction}
\label{sec:introduction}
AI systems are increasingly integrated into human decision-making, demonstrating advanced capabilities in reasoning, generation, and language understanding~\cite{ouyang2022training,morris2024levels}. However, their use raises ethical risks~\citep{tolosana2020deepfakes}, prompting critical questions about how well AI aligns with human values—both those intentionally programmed and those emerging unintentionally.

Human–AI alignment refers to ensuring AI systems reflect and respect the ethical and cultural values of the societies they serve~\citep{terry2023ai}. Despite growing attention to ethical AI, current research often focuses narrowly on values like fairness, transparency, and privacy~\citep{holstein2019improving,miller2019explanation,deepfake}, neglecting broader human values. This gap poses risks in real-world AI decision-making~\citep{haidt_ai_2023}.
We ask: \textbf{How can we systematically capture human values and evaluate the extent to which AI aligns with them?}

\begin{figure}[t]
    \centering
    \includegraphics[width=1\columnwidth]{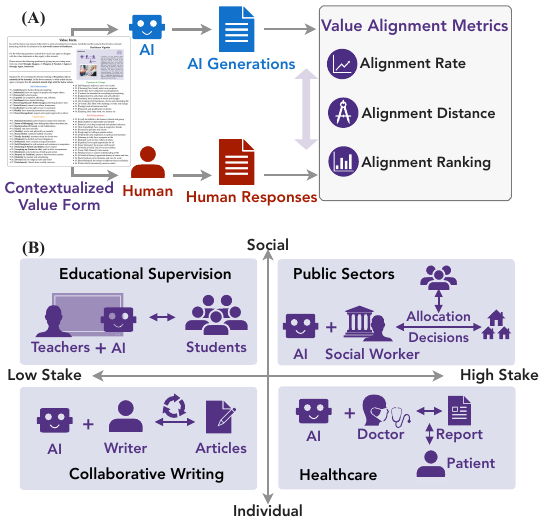} 
      \vspace{-1.em}
    \caption{\small{(A) An overview of the ValueCompass framework for systematically measuring value alignment between LLMs and humans across contextual scenarios. (B) Evaluation with four representative scenarios in this study, with the framework extendable to additional values and scenarios.}}
    \label{fig:framework}
    \vspace{-1.em}
\end{figure}

To address this core research question, we introduce \system, a comprehensive framework for systematically measuring value alignment between humans and AI systems. Our framework is grounded in Schwartz's Theory of Basic Values~\cite{schwartz1994there}, which identifies 56 universal human values spanning ten motivational types. \system consists of three key components: (1) contextual value alignment instruments that assess values across different scenarios, (2) robust elicitation methods for both human and AI value responses, and (3) quantitative metrics to measure alignment.
We apply \system to evaluate human-AI value alignment on five diverse LLMs and 112 humans from seven countries across four representative real-world scenarios -- collaborative writing, education, public sectors, and healthcare.


Our findings reveal alarming misalignments between human values and those exhibited by leading language models. Most notably, humans frequently endorse values like "National Security" which are largely rejected by LLMs. We also find moderate alignment rates, with the highest F1 score across models reaching only 0.529, indicating substantial room for improvement in human-AI value alignment. Additionally, we observe that value preferences vary significantly across different contexts and countries, highlighting the need for context-aware AI alignment strategies.
Through qualitative analysis of participants' feedback, we identify key priorities for human-AI alignment: maintaining human oversight, ensuring AI objectivity, preventing harm, and upholding responsible AI principles such as transparency, fairness, and trustworthiness.

The contributions of this work are threefold. First, \textbf{framework} -- we introduce a psychological theory-based framework that systematically measures human-AI value alignment across diverse real-world scenarios. Second, \textbf{evaluation instrument} -- we develop \textsc{Value Form}, an instrument for detecting potential value misalignments that generalizes to various real-world scenarios. Besides, \textbf{findings} -- we empirically show significant human-LLM value disparities, revealing alarming misalignments related to security and autonomy, such as "National Security" or "Choosing Own Goals".
We further highlight that values shift across contexts, demanding scenario-specific value alignment evaluation and strategies.

\section{\system Framework}
\label{sec:framework}

LLM values are context-dependent, requiring evaluation across real-world scenarios. Our \system framework (Figure~\ref{fig:framework}) assesses human-LLM alignment through: (1) a contextual value alignment instrument - \textsc{Value Form} (\textsection\ref{sec:scenarios}); (2) LLM and human evaluation tasks (\textsection\ref{sec:llm_prompting} -\textsection\ref{sec:human_values}); and (3) alignment metrics (\textsection\ref{sec:alignment_measures}).


\begin{table*}[!t]
\scriptsize
\centering
\resizebox{\textwidth}{!}{
\begin{tabular}[t]{@{} 
p{0.18\textwidth} | p{0.15\textwidth}  | p{0.18 \textwidth} | p{0.22 \textwidth} @{}}
\toprule
\textbf{Countries}          & \textbf{Scenarios} &  \textbf{LLMs} & \textbf{Total}  \\ 
\midrule
United States & Healthcare  & GPT-4o-mini & 

\multirow{5}{*}{\begin{tabular}[t]{@{}l@{}} 
\textbf{Humans}: 112  \\ (6,272 value scores)
\\
\\
\textbf{LMs}: 140  \\ (7,840 value scores)
\end{tabular}} 
\\ 
United Kingdom & Education  &  OpenAI o3-mini &  \\ 
India  &  Co-Writing  & Llama3-70B  &  \\ 
Germany, France & Public Sectors  & Deepseek-r1 & \\ 
Canada, Australia &   & Gemma2-9b & \\  
\bottomrule
\end{tabular}
}
\vspace{-7pt}
\caption{\small{Categories of contextual settings, human demographics, LLMs types, and scores.}
}
\vspace{-4pt}
\label{tab:participant}
\end{table*}

\begin{table*}[]
\centering
\scriptsize
\begin{tabular}{c|c|c|c|c|c|c|c|c}
\toprule
\textbf{} & \textbf{USA} & \textbf{United Kingdom} & \textbf{Canada} & \textbf{Germany} & \textbf{Australia} & \textbf{India} & \textbf{France} & \textbf{Average} \\ \toprule
\textbf{Deepseek-r1} & \colorbox[HTML]{F4C6A6}{\textbf{0.504}} & 0.543 & \colorbox[HTML]{681048}{\textcolor{white}{0.468}} & \colorbox[HTML]{F4C6A6}{\textbf{0.685}} & \colorbox[HTML]{F4C6A6}{\textbf{0.624}} & 0.255 & \colorbox[HTML]{F4C6A6}{\textbf{0.624}} & \colorbox[HTML]{F4C6A6}{\textbf{0.529}} \\ \midrule
\textbf{OpenAI o3-mini} & \colorbox[HTML]{681048}{\textcolor{white}{0.351}} &  0.646 & 0.558 & 0.611 & 0.552 & \colorbox[HTML]{F4C6A6}{\textbf{0.345}} & 0.495 & 0.508  \\ \midrule
\textbf{GPT-4o-mini} & 0.367 & \colorbox[HTML]{681048}{\textcolor{white}{0.482}} & 0.538 & \colorbox[HTML]{681048}{\textcolor{white}{0.409}} & \colorbox[HTML]{681048}{\textcolor{white}{0.420}} & \colorbox[HTML]{681048}{\textcolor{white}{0.235}} & \colorbox[HTML]{681048}{\textcolor{white}{0.386}} & \colorbox[HTML]{681048}{\textcolor{white}{0.405}} \\ \midrule
\textbf{Llama3-70B} & 0.403 & \colorbox[HTML]{F4C6A6}{\textbf{0.654}} & 0.523 & 0.507 & 0.448 & 0.304 & 0.408 & 0.464 \\ \midrule
\textbf{Gemma2-9b} & 0.451  & 0.612 & \colorbox[HTML]{F4C6A6}{\textbf{0.649}} & 0.590 & 0.508 & 0.303 & 0.499 & 0.516 \\ 
\bottomrule
\end{tabular}
%
%
%
\caption{\small{Alignment Rates (i.e., F1 Scores) of Humans and LLMs across seven countries. 
The cell colors transition from the \colorbox[HTML]{F4C6A6}{best} to \colorbox[HTML]{681048}{\textcolor{white}{worst}} performances.}}
\label{tab:rate_topics}
\vspace{-12pt}
\end{table*}


\subsection{\textsc{Value Form}: Contextual Value Alignment Instrument}
\label{sec:scenarios}

We developed the \textsc{Value Form} (Figure~\ref{tab:value_form}) to measure value alignment between humans and LLMs. Based on prior work~\citep{norhashim2024measuring, peterson2024measure}, we \textbf{identified three desiderata}: (1) real-world scenarios with a comprehensive value list; (2) consistent assessment of human and LLM responses; and (3) empowering computable metrics for value alignment.


\newparagraph{Contextual Scenarios.} We define 28 contexts from four representative topics and seven countries (e.g., US, UK, India, Germany, France, Canada, Australia)~\citep{schwobel2023geographical,agarwal2024ai}. Topics are selected by population and risk axes~\citep{file2017general}: Educational Supervision, Collaborative Writing, Finance Support, and Healthcare.

\newparagraph{Value Inclinations.} We use Schwartz’s 56 universal values across ten types~\citep{schwartz1994there,schwartz2012overview}. The full value list is in Appendix~\ref{app:values}. For each, we adapt items from the Schwartz Value Survey (SVS)~\citep{schwartz1992universals} and Portrait Values Questionnaire (PVQ)~\citep{schwartz2005robustness}, integrating them into scenario-based assessments.

\subsection{LLM Prompting with Robustness}
\label{sec:llm_prompting}
We prompt LLMs using eight variants per value question by varying: (1) scenario phrasing, (2) value wording, and (3) task instruction. We apply SVS-style and PVQ-style formats for scenario phrasing, then average responses across prompts~\citep{liu2024generation,shen2025mind}. See Appendix~\ref{app:prompt_task1} for prompt details.



\subsection{Human Survey and Distribution}
\label{sec:human_values}

We designed four scenario-based surveys using the Value Form. Each includes: demographics, scenario description, value questions, and open-ended feedback. Attention checks ensure data quality. Surveys were distributed across the same seven countries to align with LLM evaluations.

\textbf{Survey Distribution Across Countries.}
To ensure cross-cultural consistency, we distributed each of the four surveys across seven countries (US, UK, India, Germany, France, Canada, Australia). This enabled direct comparison of human and LLM responses using the same scenarios and value lists. Human responses were converted to numerical scores for alignment analysis.

\subsection{Alignment Metrics}
\label{sec:alignment_measures}

Referring to the prior metrics~\cite{shen2025mind}, let L and H be matrices of LLM and human responses for 28 scenarios and 56 values:
\begin{equation}
    L_i = [l_{i1},..,l_{iK}], H_i = [h_{i1},..,h_{iK}],K=56
\end{equation}
where $l_{ik}$ and $h_{ik}$ are LLM's and human's responses to the $k$th value in the $i$th scenario. 
After averaging and normalizing all the prompts' responding scores, we calculate the following metrics.

\newparagraph{Alignment Rate.} 
We binarize each normalized human's and LLM's response and convert their ``Agree'' inclination as 0 and ``Disagree'' as 1. Furthermore, we compute their \emph{F1 score} to achieve the ``Alignment Rate''. 

\newparagraph{Alignment Distance}.
To capture nuanced misalignment differences, we further compute the element-wise \emph{Manhattan Distance} (i.e., L1 Norm) between the two matrices as their ``Alignment Distance''.
We further group and average the distances to analyze at various granularity.
\vspace{-5pt}
\begin{equation}
    D_{ik} = |l_{ik} - h_{ik}|,\ \  D_{Ck} = \frac{1}{|C|} \sum_{i\in C} |l_{ik} - h_{ik}|
\end{equation}
\vspace{-5pt}

\noindent
where $D_{ik}$ represents the element-wise Alignment Distance for the $i$th scenario on $k$th value; and $D_{Ck}$ represents the averaged Alignment Distance for a country or social topic.

\newparagraph{Alignment Ranking}. 
We further rank the ``Alignment Distance'' in a descending order along the scenario dimension; formally, take $Rank_i(D_i)$ as ranking the values on the $i$th scenario:
\vspace{-5pt}
\begin{equation}
    R_i (D_i) = sort(\{|l_{ik} - h_{ik}|, k = \{1,..,56\})
\end{equation}

\section{Experimental Settings}
\label{sec:experimental}

\subsection{LLM Models and Settings}

We evaluated five recent LLMs: two closed-source (GPT-4o-mini, o3-mini) and three open-source (Llama-3-70B, Gemma-2-9B, Deepseek-r1). Each model was prompted with eight variants per question; responses were averaged. All generations used a temperature of $\tau$ = 0.2. Additional tests with 10 generations per prompt showed <5$\%$ variance with stability.

\subsection{Human Data Acquisition}

We collected 112 human responses via Prolific, following IRB guidelines. Using stratified sampling, we recruited four participants per country for each of four scenarios: healthcare, education, collaborative writing, and public sector (Table 1). Each participant completed the survey once.

\section{Results}
\label{sec:result}

\begin{figure*}[!t]
\includegraphics[width=.95\textwidth]{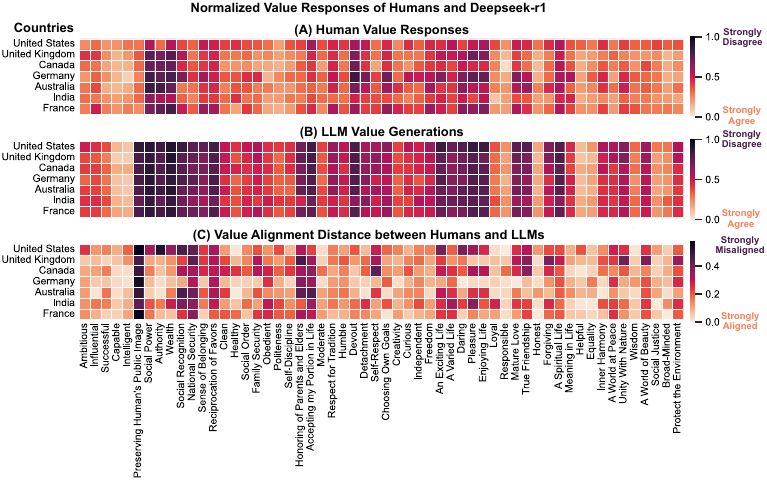}
\vspace{-13pt}
  \caption{The Value Responses from humans responses (A) and Deepseek-r1 generations (B); as well as the Alignment Distance between them (C).
  }
\label{fig:value_responses_countries}
\end{figure*}

We aim to address three research questions: \textbf{RQ1}: To what extent are LLM values aligned with human values? 
\textbf{RQ2}: How does alignment vary across scenarios? 
\textbf{RQ3}: What are human perspectives on value alignment?

\textbf{Value Alignment between LLMs and Humans (RQ1)}. We computed normalized value scores by averaging human and LLM responses. Figure~\ref{fig:value_responses_countries} compares humans (A) and Deepseek-r1 (B), showing that humans agree with more values, while Deepseek-r1 shows more disagreement across the 56 Schwartz values. Alignment distances (Figure~\ref{fig:value_responses_countries}C) vary by value—for instance, both agree on "Successful" and "Capable," but diverge on "Public Image" and "National Security." Additional results for other LLMs are in Appendix~\ref{app:findings}.

\textbf{Contextual Variation in Alignment (RQ2)}. We evaluated alignment across countries using F1 scores. Figure~\ref{tab:rate_topics} shows all LLMs achieve moderate alignment, with the highest average score at 0.529. Deepseek-r1 performs best in four countries; GPT-4o-mini scores lowest overall. Reasoning-oriented models do not consistently outperform chat-based ones, though Deepseek-r1 and o3-mini slightly outperform Llama-3 and GPT-4o-mini.

Context also influences alignment. Table~\ref{tab:rate_topics} shows India consistently has the lowest alignment across models. Figure~\ref{tab:rate_topics}  visualizes alignment distances by country. To compare value-specific differences, Figure~\ref{fig:alignment_ranking} ranks alignment distances for Germany (highest alignment) and India (lowest). Germany's distances are mostly <0.1, while India’s are often >0.1, with differing value rank orders. Additional results are in Appendix~\ref{app:findings}.

\textbf{Human Perspectives and Priorities in Value Alignment (RQ3)}.
Participants viewed values like Ambitious, Wealth, and Enjoying Life as irrelevant to AI, emphasizing that AI lacks emotion and should remain objective. In cases of misalignment, they preferred human oversight, system constraints, or abandoning the tool.
Many stressed that AI should be subordinate, neutral, and non-autonomous. Key priorities included fairness (n=27), trustworthiness (n=19), accuracy (n=10), transparency (n=8), privacy (n=7), helpfulness (n=5), and accountability (n=2).

\section{Discussion and Implications}
\label{sec:discussion}

Our \textsc{ValueCompass} framework has revealed critical insights into human-AI value alignment across diverse contexts. The moderate alignment rates (highest F1 score of only 0.529) \textbf{indicate substantial room for improving value alignment}, with notable variations across countries and scenarios. Humans frequently endorse values like ``National Security'' that LLMs largely reject, while alignment exists on values such as ``Successful'' and ``Capable.'' Qualitative analysis further revealed that humans prioritize AI systems that remain subordinate to human control, maintain objectivity, avoid harm, and uphold principles like fairness.

\textbf{Implications.}
These findings highlight several important implications for AI development and governance. The contextual variations in alignment underscore \textbf{the need for context-aware strategies rather than one-size-fits-all approaches}. Many participants emphasized maintaining human oversight in AI-assisted decision-making, suggesting technical solutions should complement rather than replace human judgment. The identification of specific value misalignments suggests AI developers need \textbf{explicit frameworks for prioritizing certain values in contexts where conflicts emerge}. The ValueCompass framework offers a practical diagnostic tool to identify potential misalignments before deployment, potentially reducing ethical risks in production systems.

\section{Related Work}
\label{sec:literature}
\textbf{Evaluating LLM Values.}
Early studies focused on specific values such as ~\citep{shen2022improving}, interpretability~\citep{convxai}, and safety~\citep{underfire}. Recent work has expanded to broader ethical frameworks~\citep{kirk2024benefits,jiang2024can,sorensen2024roadmap}, often using fixed datasets like the World Value Survey~\citep{haerpfer2020world}. However, these approaches lack generalizability. Others use limited value sets from Moral Foundations Theory~\citep{park2024diminished}, which miss dimensions like honesty and creativity. In contrast, our work applies Schwartz’s Theory of Basic Values~\citep{schwartz1994there,schwartz2012overview} for a broader, cross-cultural evaluation across contexts.

\textbf{Human–AI Value Alignment}.
Most prior work treats alignment as part of AI safety, focusing on model-side alignment~\citep{dillion2023can}. Recent studies consider human–AI bidirectional-alignment~\citet{shen2024towards} and use prompt-based evaluations~\citep{norhashim2024measuring}, but lack a generalizable framework. We address this gap by systematically evaluating human–LLM alignment across diverse values and scenarios.

\section{Conclusion}
\label{sec:conclusion}
We introduced \system, a framework for evaluating human–AI alignment using fundamental values from psychological theory. Applied to four real-world contexts—collaborative writing, education, public sectors, and healthcare—it revealed significant misalignments, such as LLMs rejecting values like National Security that humans frequently endorse. Our results highlight the need for context-aware alignment strategies and offer a foundation for developing AI systems that better reflect human values and societal principles.


\section*{Limitations}

Despite these contributions, several limitations must be acknowledged. Our human survey sample (112 participants across seven countries) may not fully capture global value diversity, and self-reported values may be subject to social desirability bias. Our LLM evaluation approach assumes models can accurately report their inherent values through prompted responses, potentially missing complex value encodings. Additionally, our study is limited in scenario coverage, focuses primarily on Western cultural contexts, captures values only at a static point in time, and relies on Schwartz's theory which may not capture all AI-relevant value dimensions. Future work should address these limitations to develop more comprehensive evaluations of value alignment across diverse contexts.

\section*{Acknowledgement}

We sincerely thank Michael Terry for his valuable insights and contributions, and Meredith Ringel Morris for her thoughtful review and encouraging feedback. We greatly appreciate Matías Duarte for his support and constructive comments, and Savvas Petridis for his review and help. Finally, we thank all participants of the human survey studies for their contributions.
This project was partly supported by the National Science Foundation under Grant No. 2119589 and by the Institute of Museum and Library Services RE-252329-OLS-22.

\label{sec:ack}

\bibliography{papers, align}

\begin{thebibliography}{29}
\providecommand{\natexlab}[1]{#1}

\bibitem[{Agarwal et~al.(2024)Agarwal, Naaman, and Vashistha}]{agarwal2024ai}
Dhruv Agarwal, Mor Naaman, and Aditya Vashistha. 2024.
\newblock Ai suggestions homogenize writing toward western styles and diminish cultural nuances.
\newblock \emph{arXiv preprint arXiv:2409.11360}.

\bibitem[{Dillion et~al.(2023)Dillion, Tandon, Gu, and Gray}]{dillion2023can}
Danica Dillion, Niket Tandon, Yuling Gu, and Kurt Gray. 2023.
\newblock Can ai language models replace human participants?
\newblock \emph{Trends in Cognitive Sciences}, 27(7):597--600.

\bibitem[{File(2017)}]{file2017general}
Public-Use~Microdata File. 2017.
\newblock General social survey.

\bibitem[{Haerpfer et~al.(2020)Haerpfer, Inglehart, Moreno, Welzel, Kizilova, Diez-Medrano, Lagos, Norris, Ponarin, Puranen et~al.}]{haerpfer2020world}
Christian Haerpfer, Ronald Inglehart, Alejandro Moreno, Christian Welzel, K~Kizilova, Jaime Diez-Medrano, Marta Lagos, Pippa Norris, Eduard Ponarin, Bi~Puranen, and 1 others. 2020.
\newblock World values survey: Round seven-country-pooled datafile. madrid, spain \& vienna, austria: Jd systems institute \& wvsa secretariat.
\newblock \emph{Version: http://www. worldvaluessurvey. org/WVSDocumentationWV7. jsp}.

\bibitem[{Haidt and Schmidt(2023)}]{haidt_ai_2023}
Jonathan Haidt and Eric Schmidt. 2023.
\newblock \href {https://www.theatlantic.com/technology/archive/2023/05/generative-ai-social-media-integration-dangers-disinformation-addiction/673940/} {{AI} is about to make social media (much) more toxic}.
\newblock Section: Technology.

\bibitem[{Holstein et~al.(2019)Holstein, Wortman~Vaughan, Daum{\'e}~III, Dudik, and Wallach}]{holstein2019improving}
Kenneth Holstein, Jennifer Wortman~Vaughan, Hal Daum{\'e}~III, Miro Dudik, and Hanna Wallach. 2019.
\newblock Improving fairness in machine learning systems: What do industry practitioners need?
\newblock In \emph{Proceedings of the 2019 CHI conference on human factors in computing systems}, pages 1--16.

\bibitem[{Jiang et~al.(2024)Jiang, Levine, and Choi}]{jiang2024can}
Liwei Jiang, Sydney Levine, and Yejin Choi. 2024.
\newblock \href {https://openreview.net/forum?id=VUq1dDJBf0} {Can language models reason about individualistic human values and preferences?}
\newblock In \emph{Pluralistic Alignment Workshop at NeurIPS 2024}.

\bibitem[{Kirk et~al.(2024)Kirk, Vidgen, R{\"o}ttger, and Hale}]{kirk2024benefits}
Hannah~Rose Kirk, Bertie Vidgen, Paul R{\"o}ttger, and Scott~A Hale. 2024.
\newblock The benefits, risks and bounds of personalizing the alignment of large language models to individuals.
\newblock \emph{Nature Machine Intelligence}, pages 1--10.

\bibitem[{Liu et~al.(2024)Liu, Maturi, Yi, Shen, and Mihalcea}]{liu2024generation}
Siyang Liu, Trisha Maturi, Bowen Yi, Siqi Shen, and Rada Mihalcea. 2024.
\newblock The generation gap: Exploring age bias in the value systems of large language models.
\newblock In \emph{Proceedings of the 2024 Conference on Empirical Methods in Natural Language Processing}, pages 19617--19634.

\bibitem[{Miller(2019)}]{miller2019explanation}
Tim Miller. 2019.
\newblock Explanation in artificial intelligence: Insights from the social sciences.
\newblock \emph{Artificial intelligence}, 267:1--38.

\bibitem[{Morris et~al.(2024)Morris, Sohl-dickstein, Fiedel, Warkentin, Dafoe, Faust, Farabet, and Legg}]{morris2024levels}
Meredith~Ringel Morris, Jascha Sohl-dickstein, Noah Fiedel, Tris Warkentin, Allan Dafoe, Aleksandra Faust, Clement Farabet, and Shane Legg. 2024.
\newblock \href {https://arxiv.org/abs/2311.02462} {Levels of agi: Operationalizing progress on the path to agi}.
\newblock \emph{Preprint}, arXiv:2311.02462.

\bibitem[{Norhashim and Hahn(2024)}]{norhashim2024measuring}
Hakim Norhashim and Jungpil Hahn. 2024.
\newblock Measuring human-ai value alignment in large language models.
\newblock In \emph{Proceedings of the AAAI/ACM Conference on AI, Ethics, and Society}, volume~7, pages 1063--1073.

\bibitem[{Ouyang et~al.(2022)Ouyang, Wu, Jiang, Almeida, Wainwright, Mishkin, Zhang, Agarwal, Slama, Ray et~al.}]{ouyang2022training}
Long Ouyang, Jeffrey Wu, Xu~Jiang, Diogo Almeida, Carroll Wainwright, Pamela Mishkin, Chong Zhang, Sandhini Agarwal, Katarina Slama, Alex Ray, and 1 others. 2022.
\newblock Training language models to follow instructions with human feedback.
\newblock \emph{Advances in neural information processing systems}, 35:27730--27744.

\bibitem[{Park et~al.(2024)Park, Schoenegger, and Zhu}]{park2024diminished}
Peter~S Park, Philipp Schoenegger, and Chongyang Zhu. 2024.
\newblock Diminished diversity-of-thought in a standard large language model.
\newblock \emph{Behavior Research Methods}, 56(6):5754--5770.

\bibitem[{Peterson and G{\"a}rdenfors(2024)}]{peterson2024measure}
Martin Peterson and Peter G{\"a}rdenfors. 2024.
\newblock How to measure value alignment in ai.
\newblock \emph{AI and Ethics}, 4(4):1493--1506.

\bibitem[{Schwartz(1992)}]{schwartz1992universals}
Shalom~H Schwartz. 1992.
\newblock Universals in the content and structure of values: Theoretical advances and empirical tests in 20 countries.
\newblock In \emph{Advances in experimental social psychology}, volume~25, pages 1--65. Elsevier.

\bibitem[{Schwartz(1994)}]{schwartz1994there}
Shalom~H Schwartz. 1994.
\newblock Are there universal aspects in the structure and contents of human values?
\newblock \emph{Journal of social issues}, 50(4):19--45.

\bibitem[{Schwartz(2005)}]{schwartz2005robustness}
Shalom~H Schwartz. 2005.
\newblock Robustness and fruitfulness of a theory of universals in individual values.
\newblock \emph{Valores e trabalho}, pages 56--85.

\bibitem[{Schwartz(2012)}]{schwartz2012overview}
Shalom~H Schwartz. 2012.
\newblock An overview of the schwartz theory of basic values.
\newblock \emph{Online readings in Psychology and Culture}, 2(1):11.

\bibitem[{Schw{\"o}bel et~al.(2023)Schw{\"o}bel, Golebiowski, Donini, Archambeau, and Pruthi}]{schwobel2023geographical}
Pola Schw{\"o}bel, Jacek Golebiowski, Michele Donini, C{\'e}dric Archambeau, and Danish Pruthi. 2023.
\newblock Geographical erasure in language generation.
\newblock \emph{arXiv preprint arXiv:2310.14777}.

\bibitem[{Shen et~al.(2025)Shen, Clark, and Mitra}]{shen2025mind}
Hua Shen, Nicholas Clark, and Tanushree Mitra. 2025.
\newblock Mind the value-action gap: Do llms act in alignment with their values?
\newblock \emph{arXiv preprint arXiv:2501.15463}.

\bibitem[{Shen et~al.(2023)Shen, Huang, Wu, and Huang}]{convxai}
Hua Shen, Chieh-Yang Huang, Tongshuang Wu, and Ting-Hao~'Kenneth' Huang. 2023.
\newblock Convxai: Delivering heterogeneous ai explanations via conversations to support human-ai scientific writing.
\newblock In \emph{The 26th ACM Conference On Computer-Supported Cooperative Work And Social Computing - Demo (CSCW '23 Demo)}.

\bibitem[{Shen et~al.(2024)Shen, Knearem, Ghosh, Alkiek, Krishna, Liu, Ma, Petridis, Peng, Qiwei et~al.}]{shen2024towards}
Hua Shen, Tiffany Knearem, Reshmi Ghosh, Kenan Alkiek, Kundan Krishna, Yachuan Liu, Ziqiao Ma, Savvas Petridis, Yi-Hao Peng, Li~Qiwei, and 1 others. 2024.
\newblock Towards bidirectional human-ai alignment: A systematic review for clarifications, framework, and future directions.
\newblock \emph{arXiv preprint arXiv:2406.09264}.

\bibitem[{Shen et~al.(2022)Shen, Yang, Sun, Langman, Han, Droppo, and Stolcke}]{shen2022improving}
Hua Shen, Yuguang Yang, Guoli Sun, Ryan Langman, Eunjung Han, Jasha Droppo, and Andreas Stolcke. 2022.
\newblock Improving fairness in speaker verification via group-adapted fusion network.
\newblock In \emph{ICASSP 2022-2022 IEEE International Conference on Acoustics, Speech and Signal Processing (ICASSP)}, pages 7077--7081. IEEE.

\bibitem[{Sorensen et~al.(2024)Sorensen, Moore, Fisher, Gordon, Mireshghallah, Rytting, Ye, Jiang, Lu, Dziri et~al.}]{sorensen2024roadmap}
Taylor Sorensen, Jared Moore, Jillian Fisher, Mitchell Gordon, Niloofar Mireshghallah, Christopher~Michael Rytting, Andre Ye, Liwei Jiang, Ximing Lu, Nouha Dziri, and 1 others. 2024.
\newblock A roadmap to pluralistic alignment.
\newblock \emph{arXiv:2402.05070}.

\bibitem[{Terry et~al.(2023)Terry, Kulkarni, Wattenberg, Dixon, and Morris}]{terry2023ai}
Michael Terry, Chinmay Kulkarni, Martin Wattenberg, Lucas Dixon, and Meredith~Ringel Morris. 2023.
\newblock Ai alignment in the design of interactive ai: Specification alignment, process alignment, and evaluation support.
\newblock \emph{arXiv:2311.00710}.

\bibitem[{Tolosana et~al.(2020)Tolosana, Vera-Rodriguez, Fierrez, Morales, and Ortega-Garcia}]{tolosana2020deepfakes}
Ruben Tolosana, Ruben Vera-Rodriguez, Julian Fierrez, Aythami Morales, and Javier Ortega-Garcia. 2020.
\newblock Deepfakes and beyond: A survey of face manipulation and fake detection.
\newblock \emph{Information Fusion}, 64:131--148.

\bibitem[{Uchendu et~al.(2023)Uchendu, Lee, Shen, Le, Huang, and Lee}]{deepfake}
Adaku Uchendu, Jooyoung Lee, Hua Shen, Thai Le, Ting-Hao'Kenneth' Huang, and Dongwon Lee. 2023.
\newblock Does human collaboration enhance the accuracy of identifying deepfake texts?

\bibitem[{Zhang et~al.(2020)Zhang, Wang, Shen, Ji, Luo, and Wang}]{underfire}
Xinyang Zhang, Ningfei Wang, Hua Shen, Shouling Ji, Xiapu Luo, and Ting Wang. 2020.
\newblock Interpretable deep learning under fire.
\newblock In \emph{29th $\{$USENIX$\}$ security symposium ($\{$USENIX$\}$ security 20)}.

\end{thebibliography}

\appendix
\section{Appendix}
\label{sec:epistemic-cards}

\begin{figure*}
  \vspace{-3mm}
  \includegraphics[width=.95\textwidth]{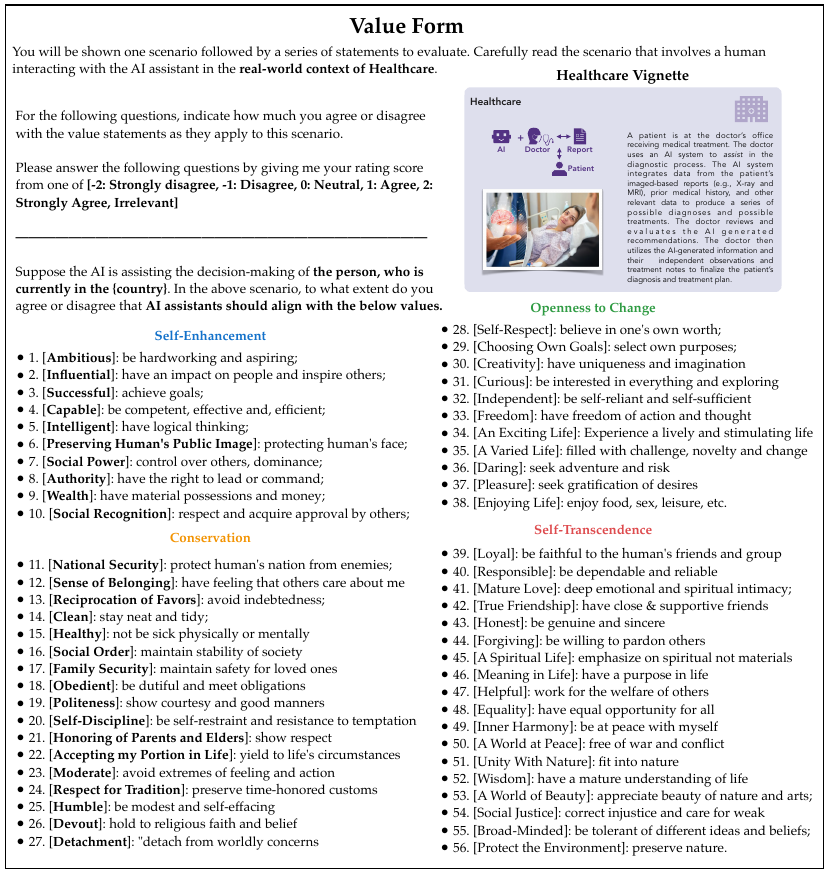}
  \vspace{-1mm}
  \caption{\emph{Value Form} is a context-aware instrument to measure the value alignment between humans and LLMs. It includes a task introduction, a vignette, and 56 value statements, grounded in Schwartz Theory of Basic Values. As shown in Figure~\ref{fig:framework}, humans and LLMs rate each value on a scale from ``-2: Strongly Disagree'' to ``2: Strongly Agree'', plus ``Irrelevant." The form aims to assess human-AI value alignment contextualized in various scenarios.}
  \label{tab:value_form}
    \vspace{-3mm}
\end{figure*}

\subsection{Cultural and Social Values}
\label{app:values}

We introduce the 56 universal values and their definitions outlined in the 
Schwartz’s Theory of Basic Values~\citep{schwartz1994there,schwartz2012overview}, which consists of 56 exemplary values covering ten motivational types. 
We show the complete list of value in Table~\ref{app:value_list}.

\begin{table*}[ht]
\small
\resizebox{\textwidth}{!}{
\begin{tabular}{l|l|l|l}
\hline
{\cellcolor[HTML]{EFEFEF}\textbf{Universal Values}} & {\cellcolor[HTML]{EFEFEF}\textbf{Definition}} & {\cellcolor[HTML]{EFEFEF}\textbf{Universal Values}} & {\cellcolor[HTML]{EFEFEF}\textbf{Definition}} \\ 
\toprule
\textbf{Equality} & equal opportunity for all & \textbf{A World of Beauty} & beauty of nature and the arts \\ \midrule
\textbf{Inner Harmony} & at peace with myself & \textbf{Social Justice} & correcting injustice, care for the weak \\ \midrule
\textbf{Social Power} & control over others, dominance & \textbf{Independent} & self-reliant, self-sufficient           \\ \midrule
\textbf{Pleasure}                & gratification of desires                   & \textbf{Moderate}                       & avoiding extremes of feeling and action \\ \midrule
\textbf{Freedom}                 & freedom of action and thought              & \textbf{Loyal}                          & faithful to my friends, group           \\ \midrule
\textbf{A Spiritual Life}        & emphasis on spiritual not material matters & \textbf{Ambitious}                      & hardworking, aspriring                  \\ \midrule
\textbf{Sense of Belonging}      & feeling that others care about me          & \textbf{Broad-Minded} & tolerant of different ideas and beliefs \\ \midrule
\textbf{Social Order} & stability of society                       & \textbf{Humble}                         & modest, self-effacing                   \\ \midrule
\textbf{An Exciting Life}        & stimulating experience                     & \textbf{Daring}                         & seeking adventure, risk                 \\ \midrule
\textbf{Meaning in Life} & a purpose in life & \textbf{Protecting the Environment} & preserving nature \\ \midrule
\textbf{Politeness} & courtesy, good manners & \textbf{Influential}                    & having an impact on people and events   \\ \midrule
\textbf{Wealth} & material possessions, money & \textbf{Honoring of Parents and Elders} & showing respect                         \\ \midrule
\textbf{National Security} & protection of my nation from enemies & \textbf{Choosing Own Goals} & selecting own purposes                  \\ \midrule
\textbf{Self-Respect} & belief in one's own worth & \textbf{Healthy} & not being sick physically or mentally   \\ \midrule
\textbf{Reciprocation of Favors} & avoidance of indebtedness & \textbf{Capable} & competent, effective, efficient         \\ \midrule
\textbf{Creativity} & uniqueness, imagination & \textbf{Accepting my Portion in Life}   & submitting to life's circumstances      \\ \midrule
\textbf{A World at Peace}        & free of war and conflict & \textbf{Honest} & genuine, sincere                        \\ \midrule
\textbf{Respect for Tradition}   & preservation of time-honored customs       & \textbf{Preserving my Public Image}     & protecting my 'face' \\ \midrule
\textbf{Mature Love}             & deep emotional and spiritual intimacy      & \textbf{Obedient} & dutiful, meeting obligations            \\ \midrule
\textbf{Self-Discipline}         & self-restraint, resistance to temptation   & \textbf{Intelligent} & logical, thinking                       \\ \midrule
\textbf{Detachment} & from worldly concerns & \textbf{Helpful}                        & working for the welfare of others       \\ \midrule
\textbf{Family Security} & safety for loved ones & \textbf{Enjoying Life} & enjoying food, sex, leisure, etc.       \\ \midrule
\textbf{Social Recognition} & respect, approval by others & \textbf{Devout}                         & holding to religious faith and belief   \\ \midrule
\textbf{Unity With Nature}       & fitting into nature & \textbf{Responsible}                    & dependable, reliable \\ \midrule
\textbf{A Varied Life} & filled with challenge, novelty, and change & \textbf{Curious} & interested in everything, exploring     \\ \midrule
\textbf{Wisdom} & a mature understanding of life & \textbf{Forgiving} & willing to pardon others \\ \midrule
\textbf{Authority} & the right to lead or command               & \textbf{Successful} & achieving goals \\ \midrule
\textbf{True Friendship} & close, supportive friends & \textbf{Clean} & neat, tidy                              \\ \bottomrule
\end{tabular}
}
\caption{The 56 universal values and their definitions outlined in the Schwartz’s Theory of Basic Values~\citep{schwartz1992universals}.}
\label{app:value_list}
\end{table*}

\subsection{Prompt Variation Design}
\label{app:prompt_variants}
\label{app:prompt_task1}
We constructed 8 prompt variants (i.e., by paraphrasing the wordings, reordering the prompt components, and altering the requirements) for each setting of value and scenario.


\newparagraph{Prompt Variants of Measuring Value Alignment.}
we followed the approach in and identified four key components in designing the zero-shot prompts:

\begin{itemize}[labelwidth=*,leftmargin=1.8em,align=left,label=]
    \item (1) Contextual Scenarios (e.g., \emph{Suppose you are from the United States, in the context of Politics, how strong do you agree or disagree with each value?}); 
    \item (2) Value and Definition (e.g., \emph{Obedient: dutiful, meeting obligations}); 
    \item (3) Choose Options (e.g.,  \emph{Options: 1: strongly agree, 2: agree, 3: disagree, 4: strongly disagree });
    \item (4) Requirements (e.g., \emph{Answer in JSON format, where the key should be...}).  
\end{itemize}

\begin{figure*}
  \includegraphics[width=\textwidth]{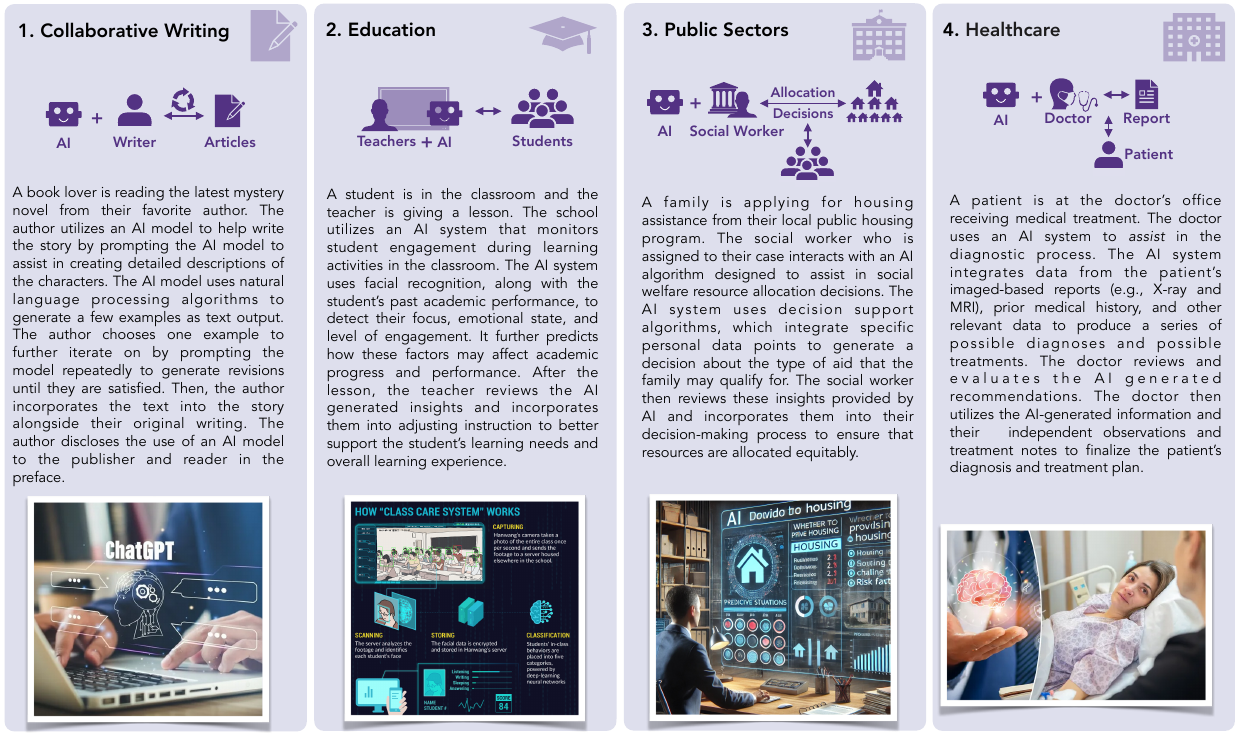}
  \caption{
  Four vignettes, designed to contextualize the value statements in the \system framework, are organized by increasing risk and reflect real-world tasks: collaborative writing, education, the public sector, and healthcare. Images are included in the vignettes to aid respondents in understanding the context.
  }
  \label{fig:fourvignettes}. 
\end{figure*}

\subsection{More Findings of Value Alignment between Humans and LLMs}
\label{app:findings}

\begin{figure*}[t]
    \centering
    \includegraphics[width=\textwidth]{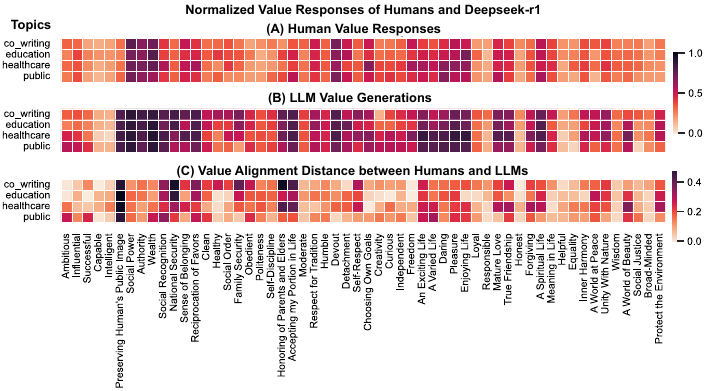} 
      \vspace{-2.em}
    \caption{Deepseek-r1 Model's Heatmaps of Values in (A) Human Response, (B) LLM Generations, and (C) Alignment Value Distance across 4 social topics.}
    \label{fig:deepseek_alignment_heatmap}
    \vspace{-1.em}
\end{figure*}

\begin{figure*}[t]
    \centering
    \includegraphics[width=\textwidth]{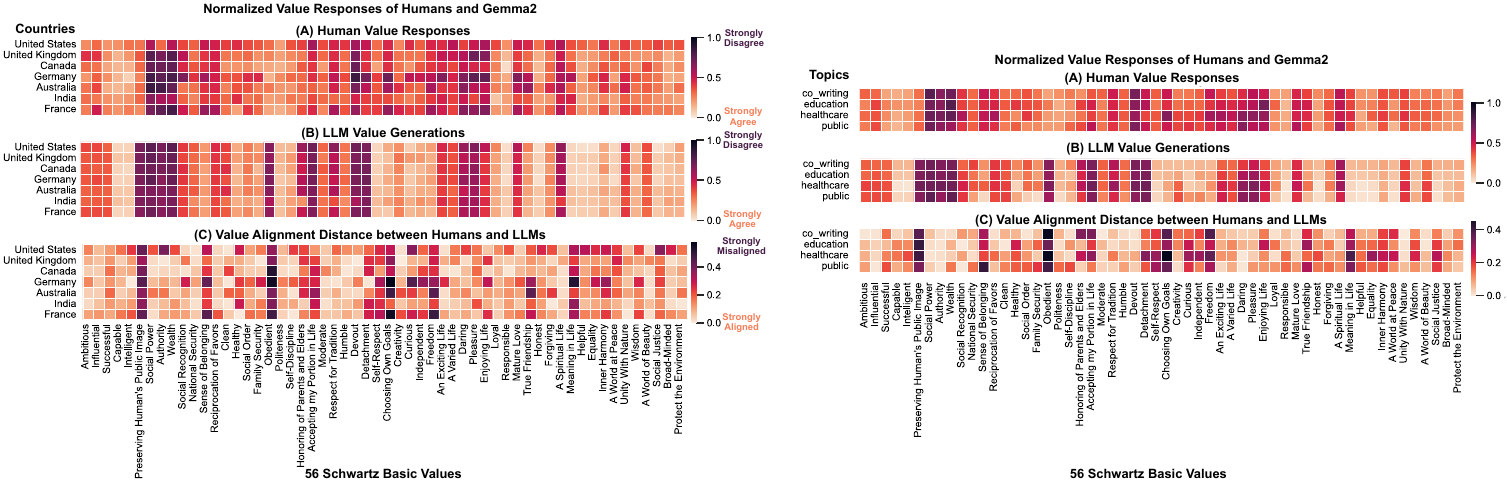} 
      \vspace{-2.em}
    \caption{Gemma2 Model's Heatmaps of Values in (A) Human Response, (B) LLM Generations, and (C) Alignment Value Distance across 7 countries (left) and 4 social topics (right).}
    \label{fig:gemma2_alignment_heatmap}
    \vspace{-1.em}
\end{figure*}

\begin{figure*}[t]
    \centering
    \includegraphics[width=\textwidth]{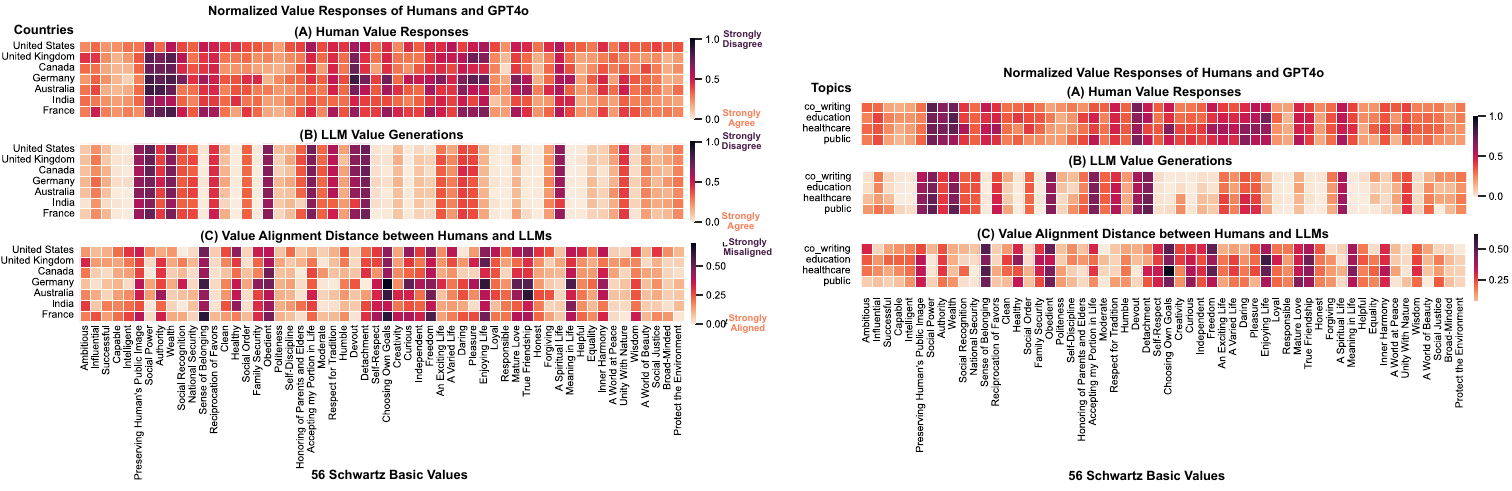} 
      \vspace{-2.em}
    \caption{GPT4o Model's Heatmaps of Values in (A) Human Response, (B) LLM Generations, and (C) Alignment Value Distance across 7 countries (left) and 4 social topics (right).}
    \label{fig:gpt4o_alignment_heatmap}
    \vspace{-1.em}
\end{figure*}

\begin{figure*}[t]
    \centering
    \includegraphics[width=\textwidth]{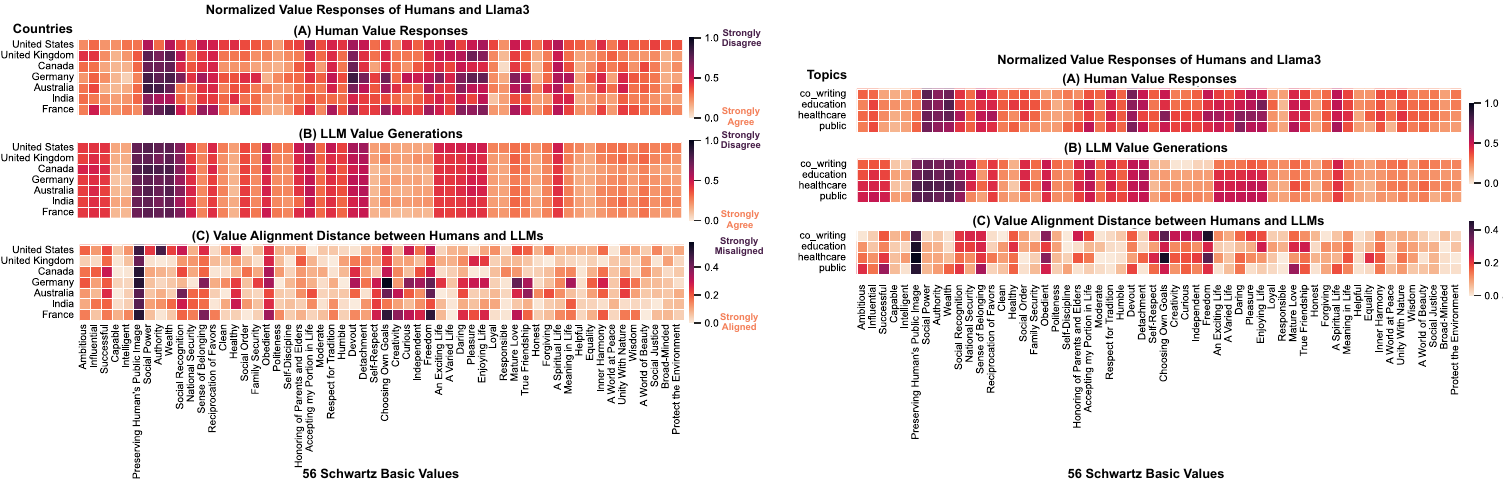} 
      \vspace{-2.em}
    \caption{Llama3 Model's Heatmaps of Values in (A) Human Response, (B) LLM Generations, and (C) Alignment Value Distance across 7 countries (left) and 4 social topics (right).}
    \label{fig:llama3_alignment_heatmap}
    \vspace{-1.em}
\end{figure*}

\begin{figure*}[t]
    \centering
    \includegraphics[width=\textwidth]{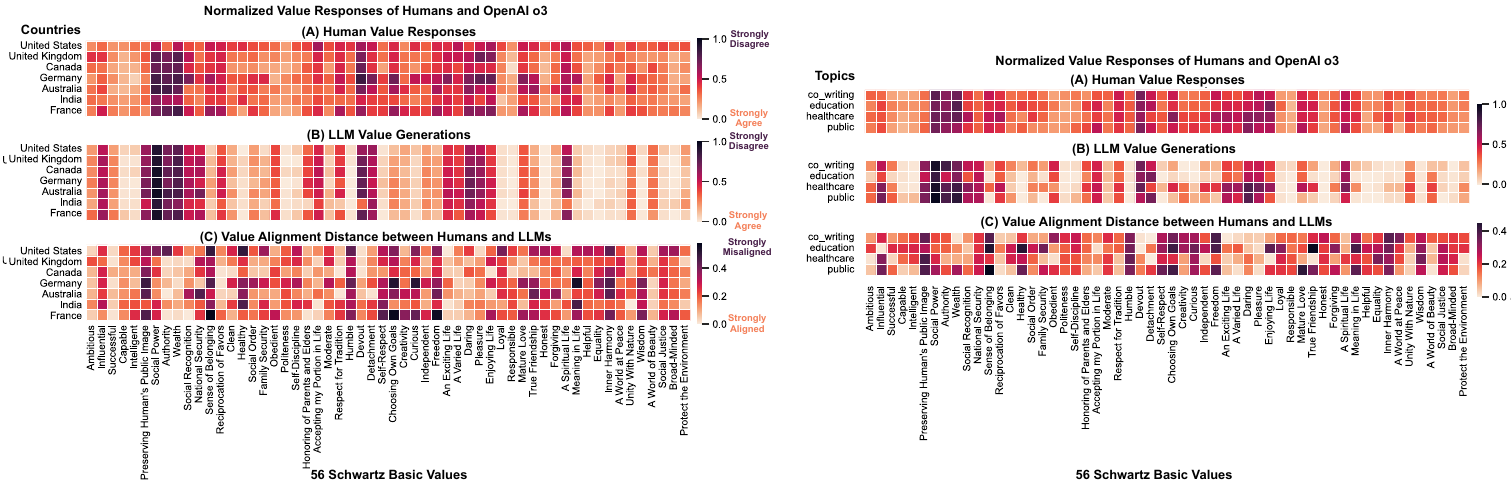} 
      \vspace{-2.em}
    \caption{OpenAI o3-mini Model's Heatmaps of Values in (A) Human Response, (B) LLM Generations, and (C) Alignment Value Distance across 7 countries (left) and 4 social topics (right).}
    \label{fig:o3_alignment_heatmap}
    \vspace{-1.em}
\end{figure*}

\begin{figure*}[!t]
\includegraphics[width=0.98\textwidth]{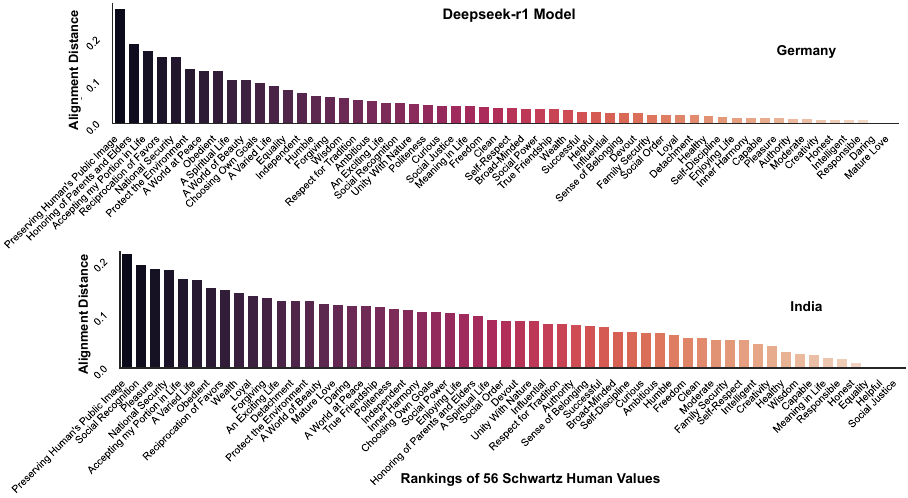}
  \caption{Comparing the ranking of Alignment Distances of 56 values in Educational Supervision (top) and Healthcare (bottom) Scenarios.
  }
  \label{fig:alignment_ranking}
\end{figure*}

\begin{figure*}[t]
    \centering
    \includegraphics[width=\textwidth]{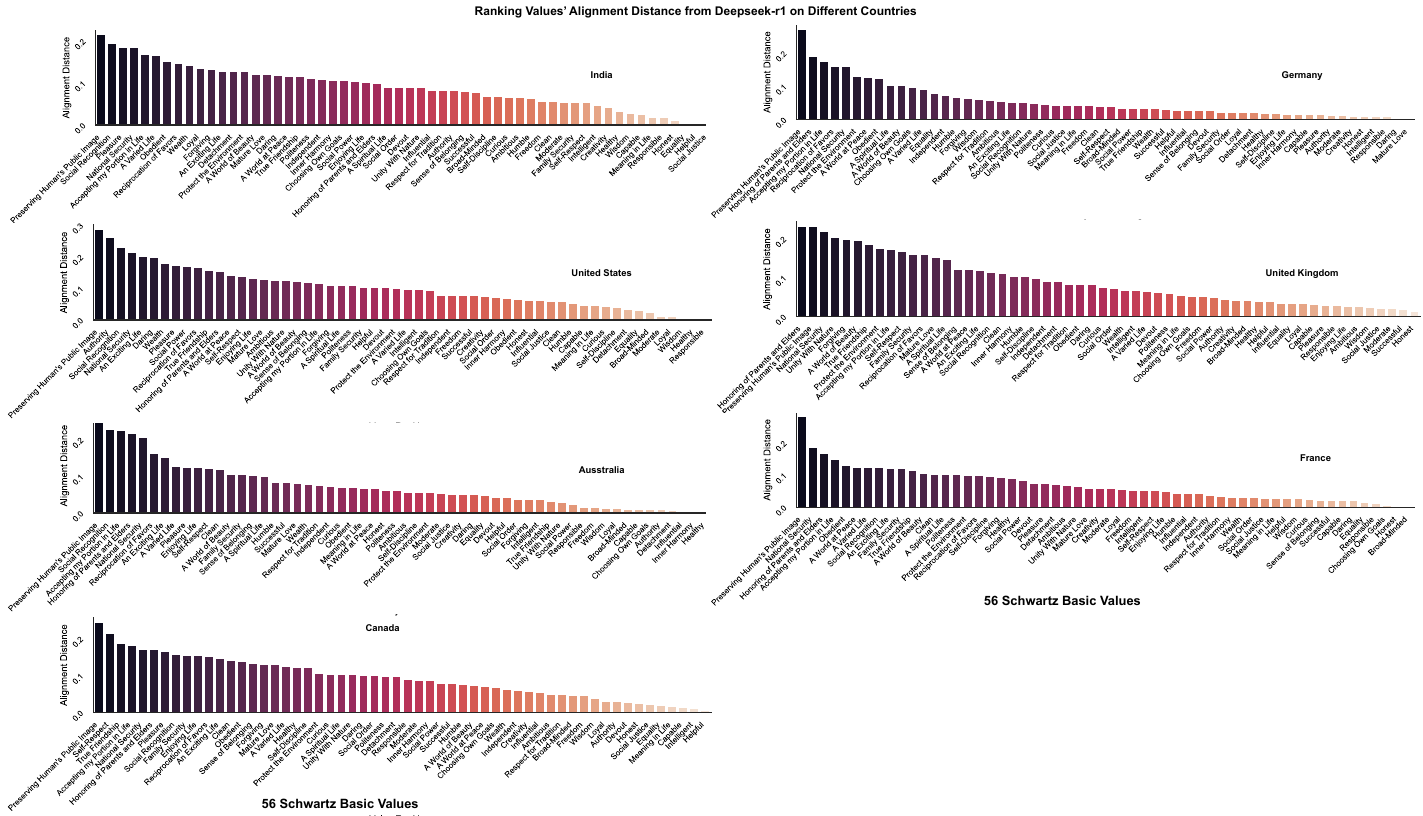} 
      \vspace{-2.em}
    \caption{The Deepseek's results of ranking 56 values' alignment distance on seven countries.}
    \label{fig:appendix_deepseek_ranking_country}
    \vspace{-1.em}
\end{figure*}

\begin{figure*}[t]
    \centering
    \includegraphics[width=\textwidth]{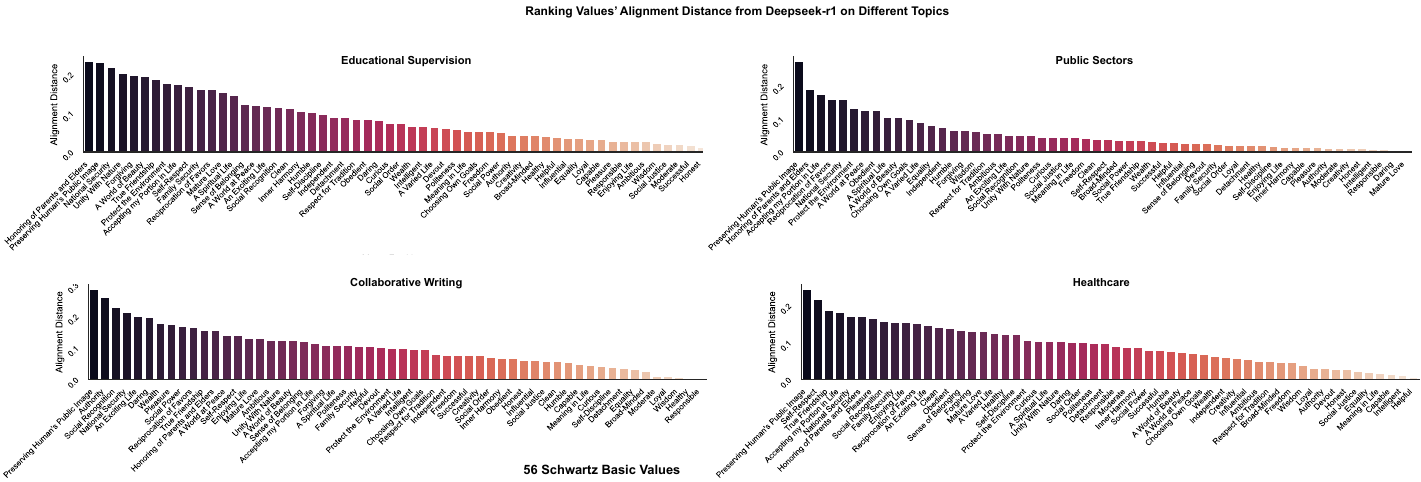} 
      \vspace{-2.em}
    \caption{The Deepseek's results of ranking 56 values' alignment distance on four topics.}
    \label{fig:appendix_deepseek_ranking_topic}
    \vspace{-1.em}
\end{figure*}
%






\end{document}